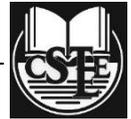

# Evaluation and Performance Analysis of the Ryu Controller in Various Network Scenarios

## Ahmadreza Montazerolghaem[1*], Somaye Imanpour[2]

*[1] Faculty of Computer Engineering, University of Isfahan, Isfahan, Iran, a.montazerolghaem@comp.ui.ac.ir*

*[2] Faculty of Computer Engineering, University of Isfahan, Isfahan, Iran, s.imanpour@eng.ui.ac.ir*

**Abstract:**

Software-defined networking (SDN) represents a revolutionary shift in network technology by decoupling the data plane from the control plane.}In this architecture, all network decision-making processes are centralized in a controller, meaning each switch receives routing information from the controller and forwards network packets accordingly. This clearly highlights the crucial role that controllers play in the overall performance of SDN. Ryu is one of the most widely used SDN controllers, known for its ease of use in research due to its support for Python programming. This makes Ryu a suitable option for experimental and academic studies. In this research, we evaluate the performance of the Ryu controller based on various network metrics and across different network topologies. For experimental analysis, we use Mininet, a powerful network emulation tool that enables the creation of diverse network structures and the connection of switches to controllers. To facilitate the experiments, we developed a Python-based script that executes various network scenarios, connects to different controllers, and captures and stores the results. This study not only provides a comprehensive performance evaluation of the Ryu controller but also paves the way for evaluating other SDN controllers in future research.

## 1. Introduction

In today's interconnected world, the role of computer networks has become indispensable. With the rapid growth in the number of applications requiring network connectivity, the limitations of traditional, rigid network architectures have become increasingly apparent.Software-defined networking (SDN) has emerged as a revolutionary technology that addresses these limitations by offering greater network flexibility, enhanced monitoring capabilities, support for custom routing algorithms, effective policy enforcement, and more efficient resource management.

For developing, testing, and evaluating algorithms, applications, and protocols designed for SDN, network emulators play a vital role. Mininet is one of the most widely used SDN emulators, enabling the creation of diverse virtual network topologies and the connection of network switches to controllers. Understanding Mininet's programming structure and running experiments through Python scripts is particularly valuable, as it provides a versatile environment for conducting a wide range of network-related research, given the limited number of studies in this domain.

In SDN architecture, all network decision-making responsibilities are centralized in the control plane, separate from the data plane. This separation underscores the importance of controllers in determining network performance and behavior. Among the available SDN controllers, Ryu [1] stands out due to its popularity and ease of use, particularly because of its support for Python-based network management and operations. This makes Ryu an ideal choice for academic research and experimentation. Evaluating and analyzing the performance of the Ryu controller can provide valuable insights and facilitate further studies in the field.

This study aims to create a robust framework for building virtual network topologies and conducting comprehensive performance evaluations of the Ryu controller. By providing a structured approach to network experimentation and performance assessment, this research contributes to both the development of SDN applications and the broader academic community.

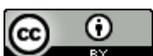

SDN not only offers a flexible and scalable alternative to traditional, complex, and inflexible networks but also opens new avenues for research. Experimentation is a crucial component of many SDN-related studies; however, existing documentation and resources for tools like Mininet and controllers like Ryu are often inadequate and time-consuming to use effectively.By developing comprehensive documentation and Python scripts for network simulation and performance evaluation, this project provides a valuable resource for researchers aiming to streamline their experimental processes.

The structure of the paper is as follows: Section 2 presents a comprehensive review of related work. Section 3 details the methodology and experimental setup used in the project. Section 4 presents the results of the experiments, visualized through graphs and analysis. Finally, Section 5 concludes the study and discusses potential directions for future research.

## 2. Related works

Extensive research has been conducted to evaluate the performance of SDN controllers, particularly the Ryu controller, which is the focus of this study. Prior works have explored various performance metrics, including throughput, delay, bandwidth, and packet loss, often using Mininet as the network emulator and a variety of traffic generation and analysis tools.

In article [2], the throughput and delay of the Ryu controller were evaluated using the Cbench tool and Mininet-based topologies. The study investigated the relationship between the number of switches, throughput, and delay, concluding that an increase in the number of switches generally results in higher delay and reduced throughput. However, the impact varies based on network topology and the running applications.

The study [3] focused on evaluating bandwidth and packet loss rates across two different topologies. The first topology included two paths with three and four switches, while the second added a third path with five switches. Using Mininet for emulation and iPerf for bandwidth measurement, the experiments were repeated five times for each configuration. Results indicated that topologies with fewer paths and without a load balancer exhibited higher bandwidth and lower packet loss. The introduction of a load balancer increased packet loss and reduced bandwidth across all configurations.

Research [4] assessed the performance of the Ryu controller by deploying a custom SDN topology in Mininet and capturing traffic using Wireshark. The study measured metrics like bandwidth, throughput, and round-trip time (RTT) for TCP and UDP traffic. Bandwidth measurements between nodes were averaged over ten runs using iPerf, while throughput and RTT were evaluated using iPerf3 and ping, respectively. The proposed custom topology demonstrated superior performance compared to the default SDN topology in terms of throughput, bandwidth, and RTT.

In article [5], a simple topology with three hosts and one switch was implemented in Mininet to analyze multiple performance metrics, including throughput, bandwidth, RTT, packet loss rate, and jitter. Using iPerf3 for TCP traffic generation, the study measured bandwidth and throughput over ten-second intervals. RTT was assessed via ICMP-based ping, while UDP-based iperf3 tests were used to evaluate packet loss and jitter across five different bandwidth levels. Results highlighted significant variations in performance based on network conditions and traffic types.

The work in [6] explored network traffic monitoring in SDN environments using the Ryu controller and three distinct Mininet topologies: a single-switch topology with 16 hosts, a linear topology with three switches and three hosts, and a tree topology with five switches and 16 hosts. Wireshark was employed for traffic observation, and key metrics like bandwidth, throughput, and jitter were evaluated. Throughput was measured over six ten-second intervals, showing maximum throughput values of 65.6 Mbps for the single-switch topology and lower values for the linear and tree topologies. Jitter analysis revealed that the tree topology managed jitter more effectively compared to other configurations.

In [7], a comparative study between the Ryu and Floodlight controllers was conducted using six different Mininet topologies. Using Qperf for testing, the study measured bandwidth and delay for each topology under both controllers. Results consistently showed that Floodlight outperformed Ryu in terms of higher bandwidth and lower delay across all topologies.

The article [8] compared the performance of Ryu and POX controllers—both Python-based—across five distinct topologies, including a single-switch topology with 100 hosts, a linear topology with 100 switches and 100 hosts, a tree topology with 127 switches and 128 hosts, a dumbbell topology with two switches and ten hosts, and a DCN topology with seven switches and eight hosts. Throughput was measured using iperf, demonstrating that Ryu achieved higher throughput across all topologies, with increases ranging from 19.88\% to 282.54\% compared to POX. Delay measurements further confirmed Ryu's superior performance, with reductions in delay between 93.48\% and 99.96\% across different configurations.

The study presented in [9] addresses the critical challenge of simultaneous load balancing in SDN-based IoMT networks by proposing an integrated approach that optimizes both server and controller load distribution. By leveraging long short-term memory (LSTM) for predictive analysis and a fuzzy system for adaptive decision-making, the proposed framework ensures efficient resource allocation across network domains. The study highlights the significance of dynamic load-balancing strategies to prevent bottlenecks at both the server and controller levels, enhancing scalability, reliability, and overall network performance. The findings contribute to improving the Quality of Service (QoS) in IoMT environments while optimizing resource utilization and reducing operational costs.

The study presented in [10] provides a comprehensive review of cloud computing and software-defined networking (SDN), focusing on architecture, resource allocation, and load-balancing strategies. It explores various algorithms, including IRMTS, reinforcement learning, and PSG, for optimizing task scheduling and network performance. Additionally, the study examines SDN's role in enhancing scalability, flexibility, and efficiency in cloud environments, while also evaluating prominent simulation tools such as Mininet and CloudSim. By analyzing existing methodologies and challenges, this research offers valuable insights into optimizing resource management and improving Quality of Service (QoS) in SDN-based cloud computing.

The study presented in [11] investigates server load balancing in multimedia Internet-of-Things (IoMT) networks using software-defined networking (SDN). By leveraging long short-term memory (LSTM) networks for load prediction and a fuzzy system for dynamic server classification, the research enhances workload distribution efficiency. The proposed method considers multiple server parameters, including CPU, memory, disk, and bandwidth utilization, to optimize resource allocation and energy consumption. This work contributes to improving load balancing strategies in SDN-based IoMT environments, addressing key challenges related to network congestion and server performance.

The study presented in [12] explores load balancing among controllers in software-defined networks (SDNs) through an improved multi-level threshold approach and switch migration operations. Unlike previous methods, which either relied on single-threshold classification or assumed uniform controller capacities, this work accurately categorizes controllers based on varying capacities to optimize switch migration decisions. By minimizing migration costs and selecting the most efficient target controllers, the proposed method enhances response time and reduces communication overhead. The findings demonstrate superior performance compared to existing schemes, contributing to more efficient and scalable SDN controller management.

The work presented in [13] introduces an optimized task scheduling approach for IoT-Fog Systems (IoTFS) leveraging Software-Defined Networking (SDN) to enhance network performance and reduce latency. By integrating a hybrid meta-heuristic algorithm, AWOA—combining Aquila Optimizer (AO) and Whale Optimization Algorithm (WOA)—the study efficiently allocates fog computing resources to IoT task requests. The proposed SDN-based AWOA method optimizes key performance metrics, including Execution Time (ET), Makespan Time (MT), and Throughput Time (TT), demonstrating superior efficiency compared to existing approaches in terms of Quality of Service (QoS) improvements.

The study presented in [14] investigates server load balancing in software-defined multimedia IoT networks by integrating Long Short-Term Memory (LSTM) for load prediction and a fuzzy system for dynamic load categorization. By utilizing four key server metrics—CPU, memory, disk, and bandwidth—the proposed approach enhances load distribution efficiency. The results demonstrate that LSTM's predictive accuracy, combined with a four-level fuzzy classification system, significantly improves resource allocation, reduces congestion, and optimizes network performance. This research provides valuable insights into adaptive load balancing strategies, offering a more efficient framework for managing multimedia traffic in IoT environments.

These studies collectively underscore the importance of evaluating SDN controller performance across diverse network topologies and traffic conditions, providing a solid foundation for further research on the Ryu controller's capabilities.

Table 1 summarizes the key metrics, tools, and topologies used in the reviewed studies:

**Table 1.** A Review of Article

| Study | Metrics Evaluated | Tools Used | Topologies Evaluated |
|-------|-------------------|------------|----------------------|
| [2] | Throughput, Delay | Cbench, Mininet | Varying switch counts |
| [3] | Bandwidth, Packet Loss | iperf, Mininet | Two paths with 3, 4, and 5 switches |
| [4] | Bandwidth, Throughput, RTT | iperf, iperf3, ping, Wireshark | Custom SDN topology |
| [5] | Throughput, Bandwidth, RTT, Packet Loss, Jitter | iperf3, ping | Three hosts, one switch |
| [6] | Bandwidth, Throughput, Jitter | Wireshark, iperf | Single-switch, Linear, Tree |
| [7] | Bandwidth, Delay | Qperf, Mininet | Six different topologies |
| [8] | Throughput, Delay | iperf, Mininet | Five distinct topologies |

## 3. Research Description

Software-defined networking (SDN) has emerged as a transformative framework that addresses the limitations of traditional network architectures, garnering substantial attention in both academic and industrial research. While developing algorithms and protocols for SDN is crucial, testing and evaluating these solutions in real-world environments are often costly and complex. Therefore, the availability of a virtual framework for evaluating SDN performance is of paramount importance. The Ryu controller, known for its Python-based programming capabilities and user-friendly documentation, has become a popular choice for academic research. This study focuses on creating various network topologies using Python scripts and the Mininet environment, conducting extensive experiments to assess network performance under different conditions.

### 3.1. Network Topologies Used in This Study

#### 3.1.1.Topology Creation Methodology

The network topologies examined in this study include various structures that are easily created using the base Topo class and related functions in Python. Each host is assigned a unique name and IP address, and the switches used are of the OVSSwitch type, supported by the Open vSwitch project. Additionally, TCLink is used as the link type, allowing the specification of parameters such as delay, bandwidth, and packet loss rate.

To create each topology, the corresponding Python script is executed. Once the topology is established in Mininet, the network nodes are verified using the 'dump' command, and the connections between them are identified with the 'links' command. The output of these commands is carefully reviewed to ensure the accurate creation of each topology. For instance, the output of these commands for a linear topology with 16 switches and 16 hosts is illustrated in the following figure. To visualize the topologies, the online tool provided by the Speer website has been utilized. In these visual representations, green circles denote hosts, gray rectangles represent switches, and the number above each switch indicates the number of directly connected hosts.

The five selected topologies include linear, star, binary tree, k-ary fat tree, and spine-leaf structures. These topologies have been implemented using generalized Python scripts to facilitate future development and expansion

#### 3.1.2.Linear (Bus) Topology

In this topology, each host is connected to a unique switch, and all switches are linearly connected to each other. This is one of the simplest possible network structures and has several limitations. However, it provides a fundamental and conceptual understanding of network behavior. As shown in Figure 1, this topology is created with 16 hosts and 16 switches.

**Figure 1.** Linear Topology with 16 Hosts and 16 Switches

#### 3.1.3.Star Topology

As depicted in Figure 2, the star topology consists of a central switch to which all other network devices are connected. Unlike the linear topology, where a single link failure may affect multiple hosts, this vulnerability is mitigated in the star topology. However, since all communications are based on the central switch, the performance of the network is heavily dependent on the efficiency of the switch and its interaction with the controller. Figure 2 illustrates this topology with five hosts.

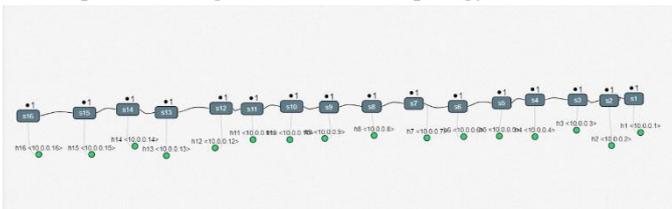

**Figure 2.** Star Topology with Five Hosts

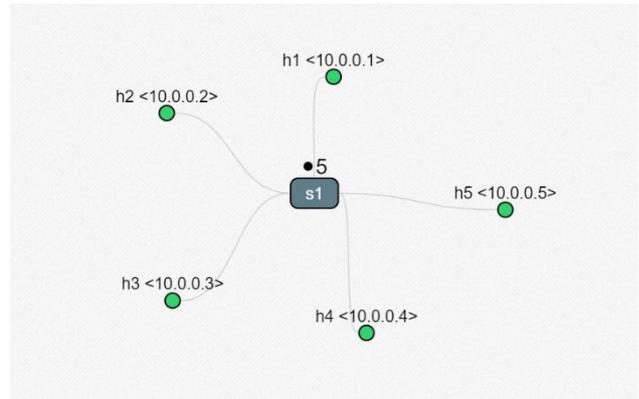

#### 3.1.4.Binary Tree Topology

In the binary tree topology, each terminal switch connects to exactly two hosts, and each non-terminal switch connects to two other switches. This hierarchical structure serves as a simple yet effective baseline for comparing other hierarchical topologies, including the Fat Tree and Spine-Leaf architectures. Figure 3 shows this topology with eight hosts.

**Figure 3.** Binary Tree Topology with Eight Hosts

#### 3.1.5.K-ary Fat Tree Topology

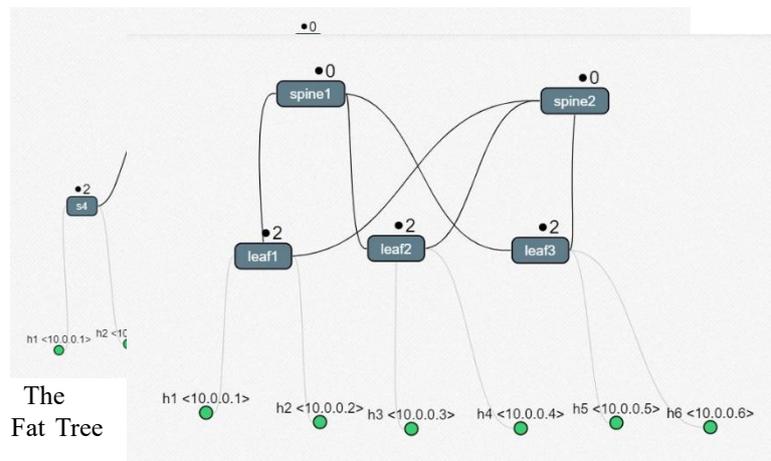

The Fat Tree topology is a three-tier network architecture consisting of core switches, aggregation switches, and edge switches. In this structure, core switches are at the highest level, connected to aggregation switches, while edge switches (also known as access switches) connect directly to the hosts. Aggregation switches act as intermediaries between edge and core switches. This topology is widely used in real-world data centers due to its load distribution capability and redundancy, providing multiple paths between switches and hosts.

The primary advantage of the Fat Tree topology lies in its ability to facilitate effective load balancing and redundancy. However, the major drawback is its high switch count, leading to increased energy consumption.

To replicate real-world network conditions, the K-ary Fat Tree was implemented in this study. This three-tier architecture features clusters, each comprising aggregation and edge switches connected to all core switches. Each cluster contains $k^2/4$ hosts, one layer of aggregation switches, and one layer of edge switches, each with k ports. Each cluster includes k/2 aggregation and k/2 edge switches, with edge switches connected to k/2 hosts and k/2 aggregation switches. Aggregation switches link to k/2 edge switches and k/2 core switches. Finally, there are $k^2/4$ core switches, each connecting to k clusters. Figure 4 illustrates the implementation of this topology for k=4.

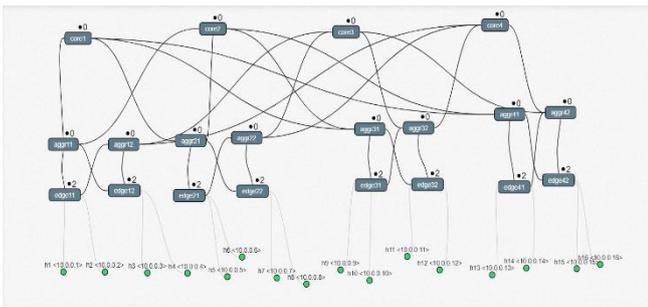

**Figure 4.** Fat Tree Structure with Four Clusters

### 3.1.6. Spine-Leaf Topology

The Spine-Leaf topology is another tree-based architecture widely used in data centers. It features two layers of switches, categorized into spine and leaf switches. Spine switches reside in the upper layer, while leaf switches occupy the lower layer and connect directly to end devices. In a full-mesh configuration, each leaf switch connects to every spine switch, enhancing load distribution and redundancy. Unlike the Fat Tree topology, the Spine-Leaf architecture requires fewer switches, reducing energy consumption.

The Python code for creating this topology is designed to generate the structure by specifying the number of spine switches, leaf switches, and the number of hosts per leaf switch. Figure 5 shows this topology with two spine switches, three leaf switches, and two hosts per leaf switch.

**Figure 5.** Spine-Leaf Structure for two spine switches, three leaf switches, and two hosts per leaf switch

**Table 2.** Summary of Network Topologies

| Topology | Number of Switches | Number of Hosts | Key Characteristics |
|---|---|---|---|
| Linear (Bus) | 16 | 16 | Simple structure, sequential connections |
| Star | 1 central switch | 5 | Centralized design, dependency on one switch |
| Binary Tree | Varies | 8 | Hierarchical structure, good baseline |
| K-ary Fat Tree | Depends on k | $k^2/4$ | High redundancy, balanced load distribution |
| Spine-Leaf | 2 spines, 3 leaves | 6 | Efficient load distribution, fewer switches |

## 3.2. Ryu Controller in the Experiments

Before conducting any experiment, the Ryu controller was initialized. In Software-Defined Networking (SDN), switches do not make any network decisions themselves; instead, all decisions are made by the controller. Given that Ryu's operation requires a Python program to define network policies, the choice of program significantly impacts the controller's performance and, consequently, the overall network behavior. In this study, two different programs were employed: the simple switch and the simple switch with Spanning Tree Protocol (STP) implementation. The following sections describe the functionality of each program.

### 3.2.1. Simple Switch Program in the Ryu Controller

Layer 2 switches gradually construct a table by learning the MAC addresses of hosts through packet forwarding. When a switch receives a packet with a known destination MAC address, it forwards the packet through the corresponding port. Otherwise, it broadcasts the packet across all ports. The simplest Ryu program for a controller implements this behavior, using Packet-In messages to learn the source address and then forwarding packets based on the table's entries or by broadcasting when necessary. While this program effectively implements a basic switch, it suffers from the broadcast storm problem, particularly in network topologies with loops.

### 3.2.2. Simple Switch with Spanning Tree Protocol Implementation in the Ryu Controller

To mitigate the broadcast storm issue, spanning tree protocols are used to create a logical loop-free topology. Various protocols exist, including the Spanning Tree Protocol (STP), Rapid Spanning Tree Protocol (RSTP), and Multiple Spanning Tree Protocol (MSTP). The implementation in Ryu used in this study is a basic version of the STP protocol. By employing this approach, network topologies with loops, such as the Fat-Tree and Leaf-Spine structures, can operate efficiently without encountering broadcast storms.

### 3.3. Experiment Setup

Given the need to repeat experiments for different network topologies and varying numbers of hosts, each experiment was implemented as a method within the Experiment class in Python. The results of each experiment were extracted and stored in CSV files to facilitate further analysis and visualization using Python libraries such as Matplotlib.

### 3.3.1. Bandwidth Measurement

To measure bandwidth, the iperf tool—pre-installed on Mininet hosts—was used. In this client-server architecture, one host acts as the client and the other as the server. The client sends data to the server, and the server measures the bandwidth and other related metrics. The sample output includes the execution time (Interval), the amount of data transferred (Transfer), and the measured bandwidth (Bandwidth).

Bandwidth tests were conducted for network topologies with 2, 4, 8, 16, 64, and 128 hosts over durations ranging from 5 to 115 seconds. The results were saved in CSV format and visualized through various plots using Matplotlib.

### 3.3.2. Throughput Calculation

Throughput was calculated using the iperf output by dividing the amount of data transferred by the transmission time. To ensure consistency in comparing throughput with bandwidth, data units were converted to match. Throughput experiments were conducted for all topologies with host counts of 2, 4, 8, 16, 64, and 128. Results were stored in CSV format and visualized using Matplotlib.

### 3.3.3. Round-Trip Time (RTT) Measurement

Round-Trip Time (RTT) was measured by using the ping command between the first and last hosts in the network.

The following steps outline the process:

1. ARP Request: The source host sends an ARP request to obtain the destination host's MAC address.

2. ARP Reply: The destination host responds with its MAC address.

3. ICMP Echo Request: Once the MAC address is known, the source host sends an ICMP echo request.

4. ICMP Echo Reply: The destination host responds with an echo reply.

Since the ARP process and its responses introduce delays, the Packet-In event occurs three times due to the initial absence of MAC address mappings. Consequently, the first ICMP message's time differs from subsequent messages and was measured separately.

For each experiment, the ping command was executed between the first and last hosts, repeating the test ten times. Network topologies such as Star, Bus, and Binary Tree were tested with 2, 4, 8, 16, 32, 64, and 128 hosts. Each test was repeated three times, and the average results were recorded as the final measurements.

## 4. Simulation analysis

In this project, using Python programming and the Mininet environment, experiments were conducted on five network topologies: linear, star, binary tree, fat tree, and spine-leaf. These experiments were conducted by evaluating three criteria—bandwidth, throughput, and delay—across each topology, using varying numbers of hosts. The results of each experiment were saved in CSV files. Using these files and the Matplotlib library, the corresponding charts for each experiment were generated. These charts help compare the experimental results between different topologies with different numbers of hosts. In this chapter, the results of each experiment are presented in the form of charts.

### 4.1. Experimental Results for Bandwidth and Throughput

As previously mentioned, bandwidth tests were conducted for all topologies. Loop-free topologies, namely star, linear, and binary tree, were tested with the simple switch program of the Ryu controller. The fat tree and spine-leaf topologies, due to the presence of loops, were tested with the spanning tree program of the Ryu controller. Given that the first measured bandwidth value in each experiment occurs when the initial routing between two hosts takes place, this value directly depends on the controller's performance and has therefore been carefully examined.

### 4.1.1. Linear Topology

The linear or bus topology was tested with the simple switch program of the Ryu controller for host counts ranging from two to 128, where the number of hosts was a power of two. Figure 6 shows the results of each experiment over different time intervals. Each value on the time axis represents the execution of the bandwidth test over the interval from zero to that value. The charts, with different colors, represent different numbers of hosts. As seen in the charts, generally, with an increase in the number of hosts and switches, bandwidth and throughput decrease. Due to the nature of this topology, as the number of hosts increases, the distance between the first and last host also increases. For the case of two hosts, there is only one link between the first and last host in addition to their unique links. For 128 hosts, there are 127 links between the first and last host. This leads to the lowest throughput observed in the final scenario. This leads to the lowest throughput observed in the final scenario. Notably, in the final experiment with 128 hosts, throughput and bandwidth remained zero until the 20-second mark. The output from Iperf for these experiments indicated a lack of routing between the first and last host with the message "connection failed: No route to host." This occurred because no communication path had been established between the first and last host during this period, due to the time needed for switch learning and interaction with the controller.

### 4.1.2.Star Topology

The star topology was tested with the Ryu controller simple switch program for host counts ranging from two to 128, where the number of hosts was a power of two. Figure 7 shows the results of each experiment over different time intervals. Each value on the time axis represents the execution of the bandwidth test over the interval from zero to that value. Different colors represent different numbers of hosts. Since each device in this topology connects to a central switch via a dedicated link—and there is only one central switch—the learning and communication process with the controller was not expected to hinder routing. As shown in the charts, the lowest bandwidth occurred during the first experiment in all scenarios. However, successful routing between the first and last host was achieved in all cases.

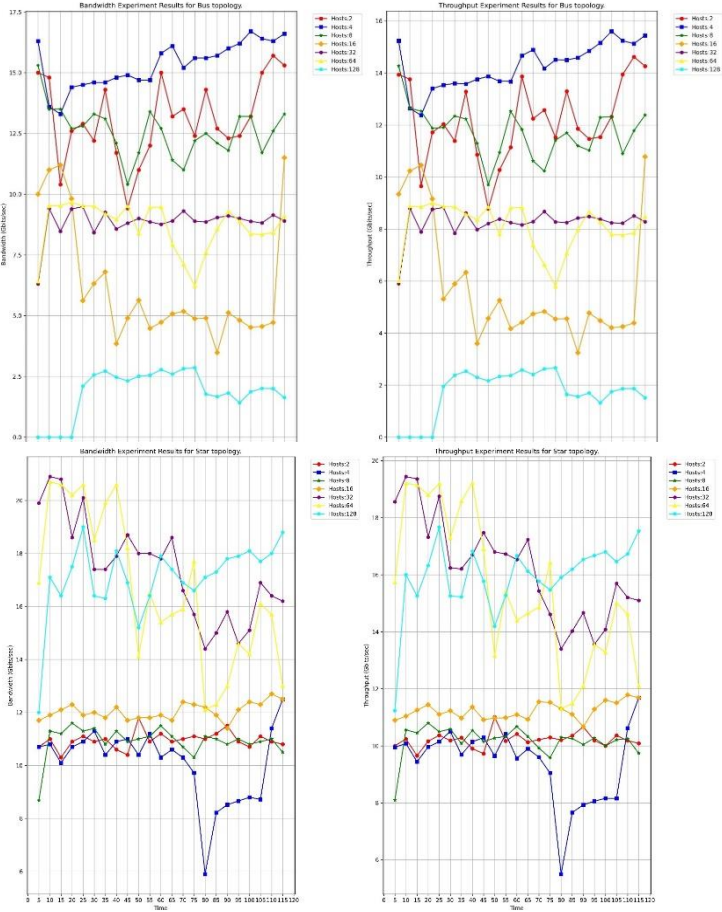

**Figure 7.** **Results of Bandwidth and Throughput Test for the Star Topology in the Time Interval from 0 to t**

### 4.1.3.Binary Tree Topology

The binary tree topology was tested with the simple switch program of the Ryu controller for host counts ranging from two to 128, where the number of hosts was a power of two. Figure 8 shows the results of each experiment over different time intervals. Each value on the time axis represents the execution of the bandwidth test over the interval from zero to that value. Different colors represent different numbers of hosts. In this topology, the lowest bandwidth also occurred during the first experiment. However, routing between the first and last host was established in under five seconds in all cases. Generally, bandwidth decreased with increasing network size in this topology.

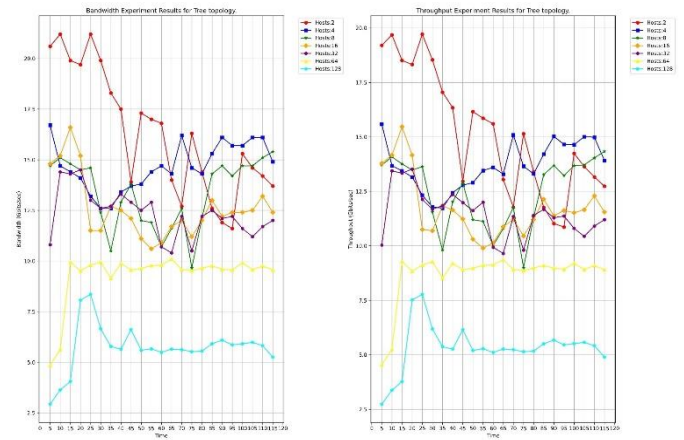

**Figure 8.** **Results of Bandwidth and Throughput Test for the Binary Tree Topology in the Time Interval from 0 to t**

### 4.1.4.Comparison of Bandwidth and Throughput in Loop-Free Topologies

To enable comparison between different network topologies with an equal number of hosts, the bandwidth charts for these topologies were drawn side by side. The results of these comparisons are shown in Figures 9 to 15. As observed, with a high number of hosts and increasing network size, the star topology exhibits higher bandwidth compared to the tree topology, and the tree topology shows higher bandwidth than the linear topology. Since each host in the star topology connects to the central switch via a unique link, it is logical that this topology maintains higher bandwidth with larger network sizes. For smaller network sizes with fewer hosts, the linear and tree topologies demonstrate similar performance. However, as the network grows, the linear topology eventually achieves higher bandwidth

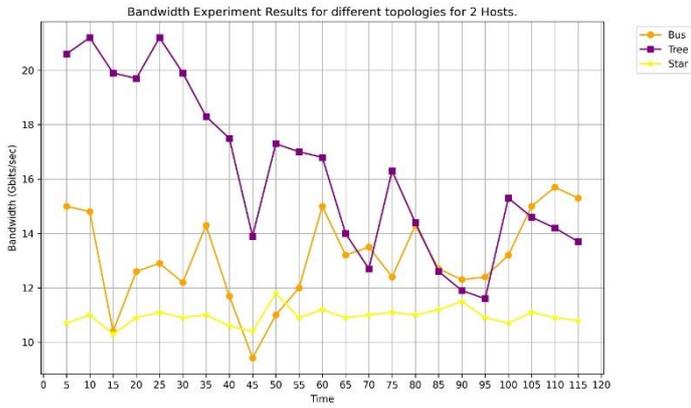

**Figure 9.** Results of Bandwidth and Throughput Test for Three Loop-Free Topologies with Two Hosts

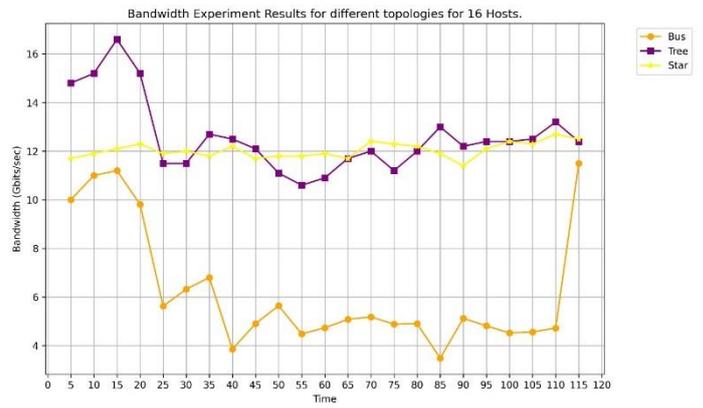

**Figure 13.** Results of Bandwidth and Throughput Test for Three Loop-Free Topologies with Thirty-Two Hosts

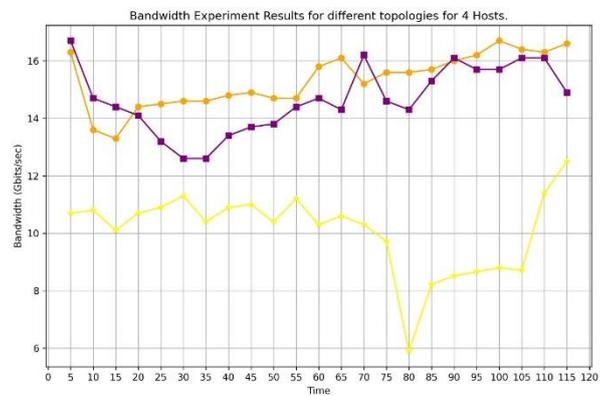

**Figure 10.** Results of Bandwidth and Throughput Test for Three Loop-Free Topologies with Four Hosts

**Figure 14.** Results of Bandwidth and Throughput Test for

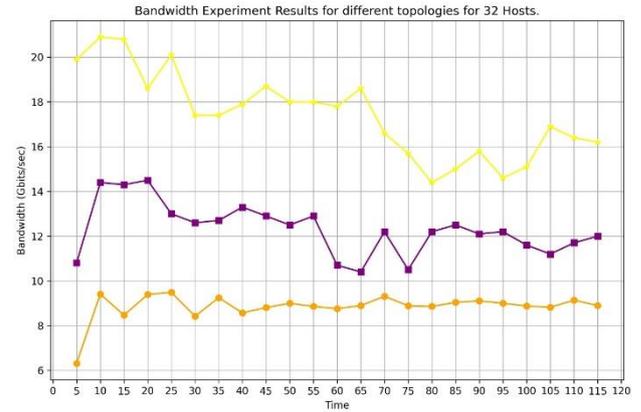

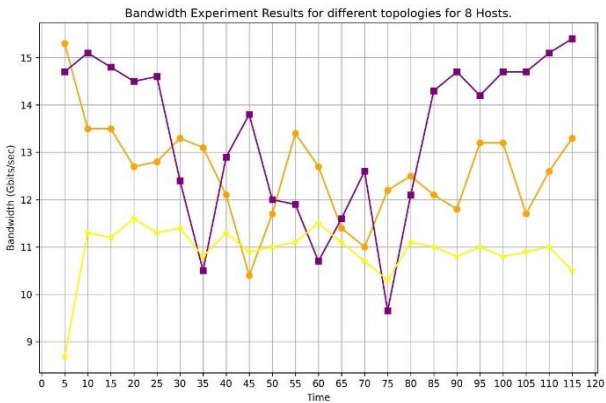

**Figure 11.** Results of Bandwidth and Throughput Test for Three Loop-Free Topologies with Eight Hosts

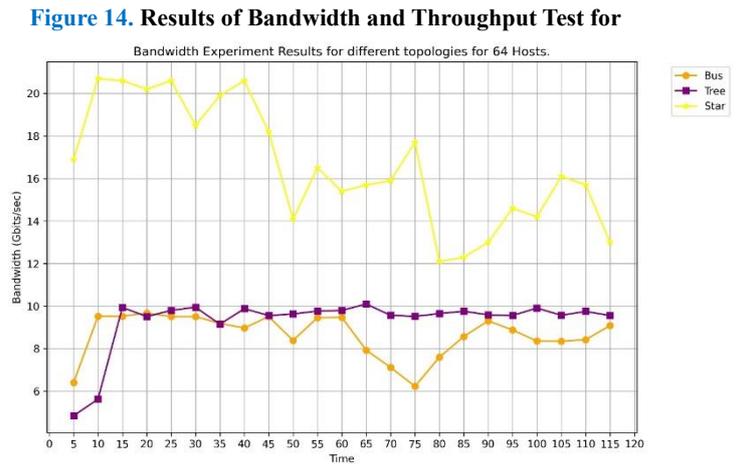

Three Loop-Free Topologies with Sixty-Four Hosts

**Figure 12.** Results of Bandwidth and Throughput Test for Three Loop-Free Topologies with Sixteen Hosts

**Figure 15.** Results of Bandwidth and Throughput Test for Three Loop-Free Topologies with One Hundred Twenty-Eight Hosts

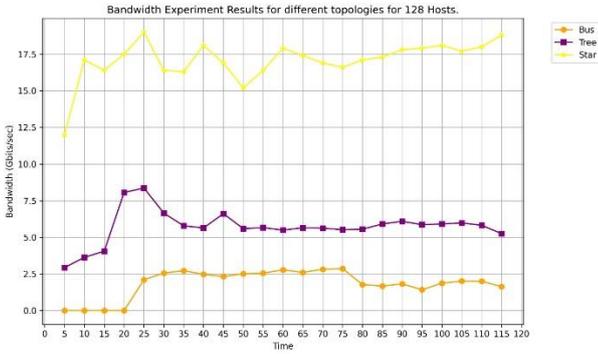

### 4.1.5. Bandwidth and Throughput Testing for Fat Tree Topology

As mentioned earlier, due to the presence of loops, the fat tree topology encounters issues when tested with the simple switch program of the Ryu controller. During the first packet transmission to the destination host, the ARP request packet is broadcasted, which causes a broadcast storm at the start of the experiment. Hence, the Ryu controller program implementing the spanning tree protocol (STP) was used for testing this topology. Prior to conducting any experiments, the controller program was allowed time to establish the spanning tree within the network structure. The fat tree topology was created with clusters ranging from two to eight, representing two to 128 hosts. Figure 16 shows the results of the bandwidth experiment for the fat tree topology. Notably, the bandwidth measured in the first experiment does not differ significantly from subsequent measurements.

**Figure 16.** Results of Bandwidth and Throughput Test for the Fat Tree Topology in the Time Interval from 0 to t

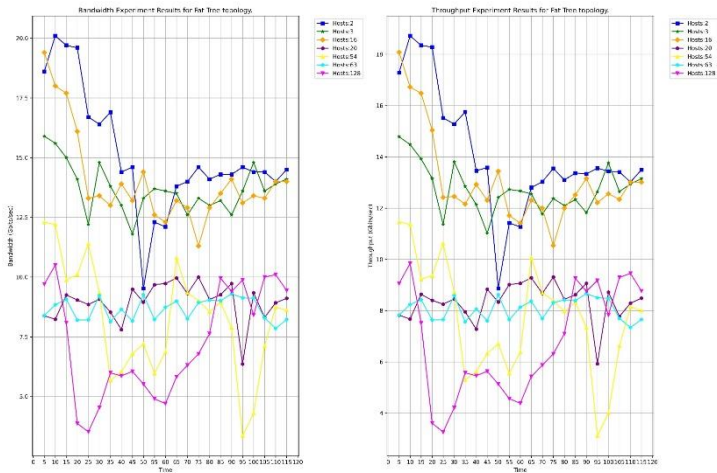

### 4.1.6. Bandwidth and Throughput Testing for Spine-Leaf Topology

The testing methodology for the spine-leaf topology was similar to the fat tree topology. Figure 17 shows the results of the bandwidth experiment for this topology. This experiment was conducted with the same number of core switches and hosts as in the fat tree topology. Notably, during the 128-host experiment, the controller encountered issues, and the test had to be repeated more than five times without success. In this topology as well, the bandwidth measured in the first experiment does not differ significantly from subsequent measurements.

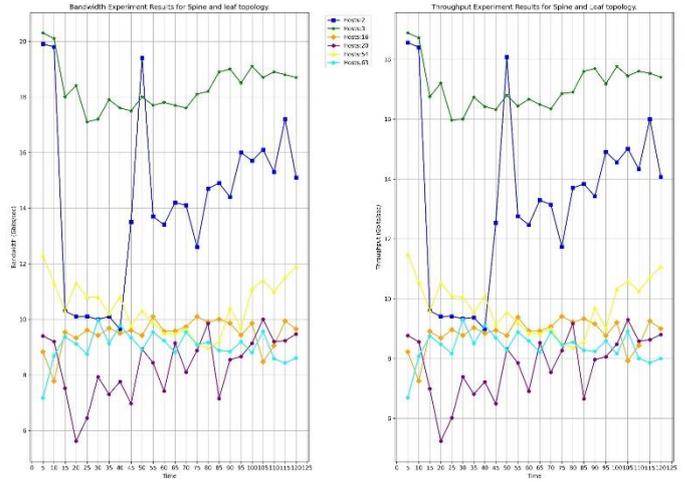

**Figure 17.** Results of Bandwidth and Throughput Test for the Spine-Leaf Topology in the Time Interval from 0 to t

### 4.1.7. Comparison of Bandwidth and Throughput in Loop-Based Topologies

Figure 18 compares the bandwidth of the fat tree and spine-leaf topologies. As shown, both topologies perform similarly for two hosts. With three hosts, the spine-leaf topology performs better, while for 16 hosts, the fat tree topology shows better performance. With a higher number of hosts, the bandwidth of the two topologies becomes very similar. Given that the spine-leaf topology uses fewer switches, it demonstrated more efficient performance in this experiment.

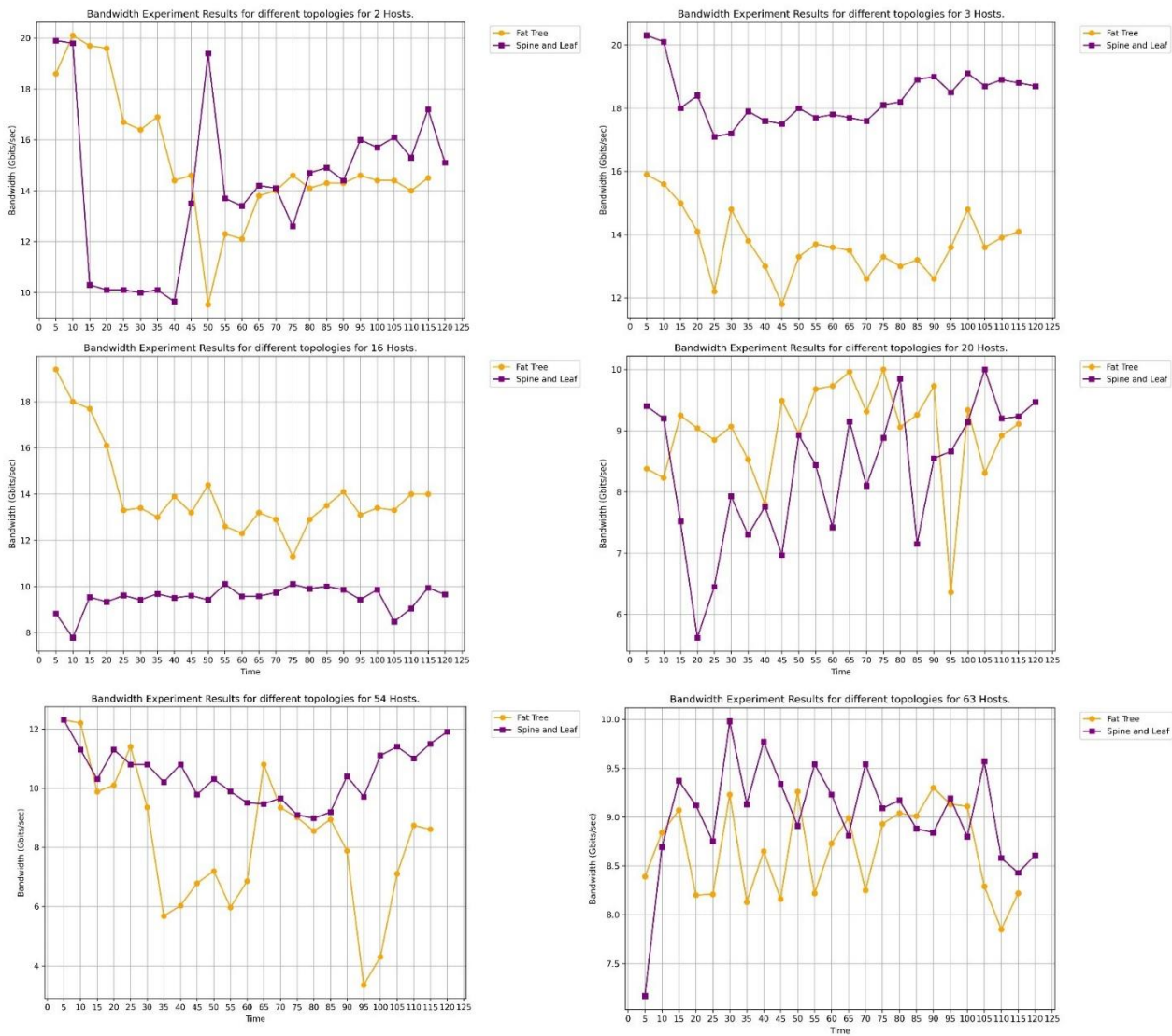

**Figure 18.** Throughput Comparison of Loop-Based Topologies

## 4.2. Experimental Results for RTT Delay

To measure delay across all topologies, the ping command was used. Given that the initial message transmission between two hosts requires the ARP protocol to discover the destination host and the switch must learn the MAC addresses of the hosts by communicating with the controller during the first message exchange, the ping command was executed only once initially for each loop-free topology, and its result was separately recorded. Subsequently, the ping command was executed ten more times, and the results were also stored. To ensure the accuracy of the experimental data, this process was repeated three times for each topology with a specified size, and the final result used in the charts represents the average of these three trials. In all charts, the RTT unit is calculated in milliseconds.

### 4.2.1.Linear Topology

Figure 19 represents the RTT results for the linear topology with varying numbers of switches. As expected, for a small number of hosts, the delay is negligible; however, as the number of hosts increases, the delay rises proportionally, reaching over 9000 milliseconds for 128 hosts. The orange line in the chart represents the delay of the first packet transmission, which is consistently higher than the maximum observed delay in all cases.

Evaluating the RTT metric in the linear topology is particularly important for the first trial because, due to the large number of switches in this structure, all switches must establish communication with the controller when the first packet is sent between two hosts. Therefore, the results of the initial experiment also reflect the controller's response time for a given number of switches.

### 4.2.2.Star Topology

Figure 20 represents the RTT results for the star topology with varying numbers of hosts. It is clearly evident from the charts that in this topology, the increase in the number of hosts does not significantly affect the delay between two specific hosts. Nonetheless, the first trial consistently shows a considerably higher delay compared to subsequent trials.

### 4.2.3.Binary Tree Topology

Figure 21 represents the RTT results for the binary tree topology. In this structure, the delay increases with the growth of the network size. The impact of network size on the first packet transmission time is more pronounced here: as the network size expands from 2 to 128 hosts, the first packet transmission delay rises from approximately 700 milliseconds to over 3000 milliseconds.

### 4.2.4. RTT Comparison in Loop-Free Topologies

Overall, the linear topology exhibits significantly higher RTT compared to the other two topologies for the same number of hosts. For large numbers of hosts, the RTT in the star and binary tree topologies is almost negligible in comparison. Figure 22 compares the delay across the loop-free topologies.

The star topology consistently maintains similar RTT values across different numbers of hosts. The binary tree topology, while exhibiting higher RTT than the star topology, shows an increasing delay as the number of hosts grows. As illustrated in Figure 23, the chart on the right presents the time taken for the first packet transmission across topologies—where the binary tree typically has the highest delay and the star topology the lowest.

### 4.2.5.Fat Tree Topology

Similar to the bandwidth experiment, this test also employed a simple switch program with the STP protocol running on the controller. Figure 24 shows the results for networks with 2, 3, 16, 20, 54, and 63 hosts.

### 4.2.6.Spine-Leaf Topology

Figure 25 presents the RTT results for the spine-leaf topology. This structure was designed to have the same number of hosts as the fat tree topology, with the number of spine switches equal to the core switches in the fat tree and the number of leaf switches matching the access switches in the fat tree.

### 4.2.7.RTT Comparison in Looped Topologies

Figure 26 comparison of the experimental results for the fat tree and spine-leaf topologies reveals that the fat tree consistently exhibits higher RTT across all cases. This result can be attributed to the greater number of switches and links in the fat tree structure for the same number of hosts tested.

This research evaluated five different network topologies with varying numbers of hosts based on bandwidth, throughput, and delay metrics using two programs from the Ryu controller. Additionally, in the throughput analysis, the first result was separately examined due to its direct interaction with the controller. In this chapter, the results of each experiment were presented through charts designed to facilitate comparisons and thorough analysis of different scenarios.

## 5. Conclusion

As discussed in the literature review, most studies evaluating the Ryu controller have been conducted on small-scale network topologies with a limited number of hosts and switches. However, evaluating Software-Defined Networks (SDN) on a scale closer to real-world applications is crucial for assessing their practical performance and reliability. In this research, we aimed to address this gap by developing code for the Mininet tool to facilitate comprehensive network testing. Mininet offers the flexibility to design various network topologies with arbitrary numbers of hosts, switches, and links. Moreover, performance evaluations of the Ryu controller were conducted across a range of network topologies—from small to relatively large scales—to achieve a more accurate and practical understanding of its capabilities.

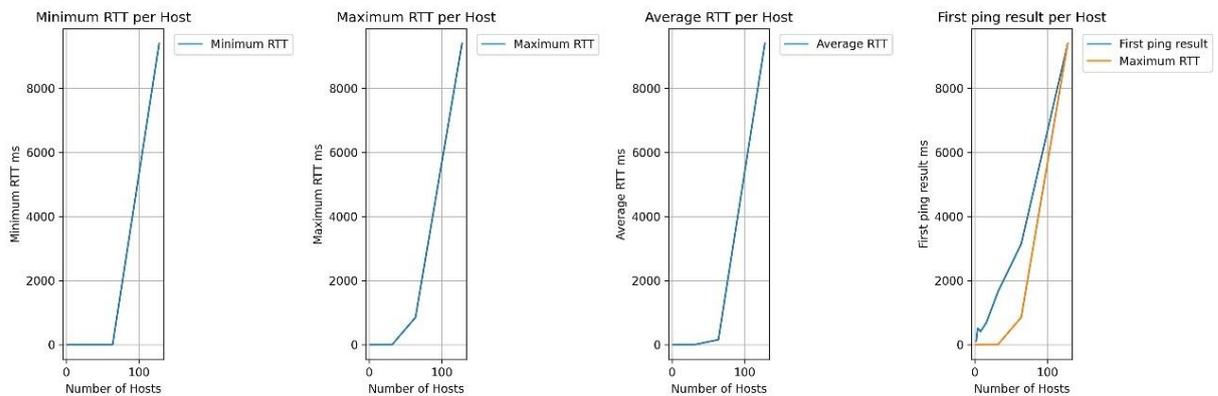

**Figure 19. RTT Test Results for the Linear Topology**

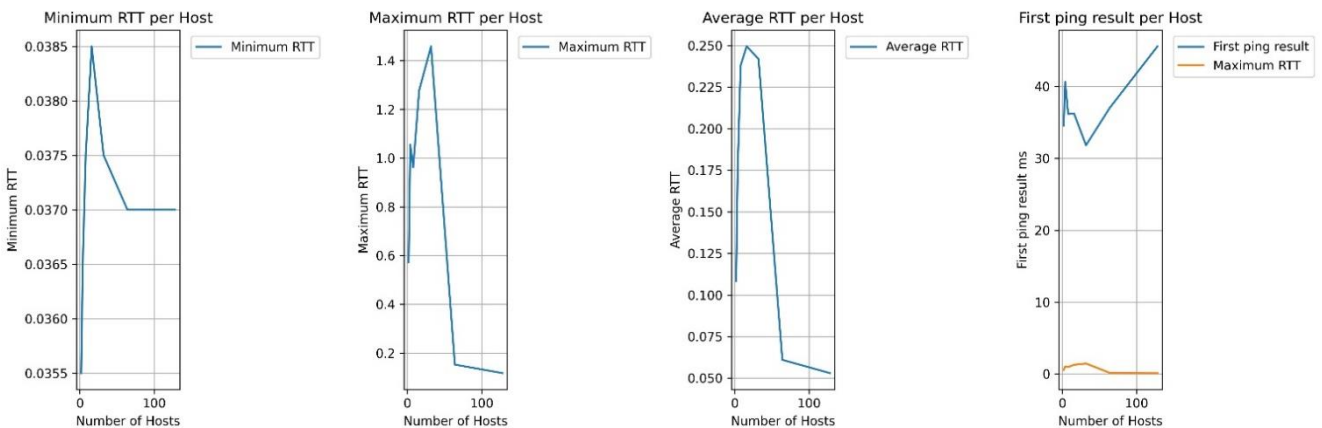

**Figure 20. RTT Test Results for the Star Topology**

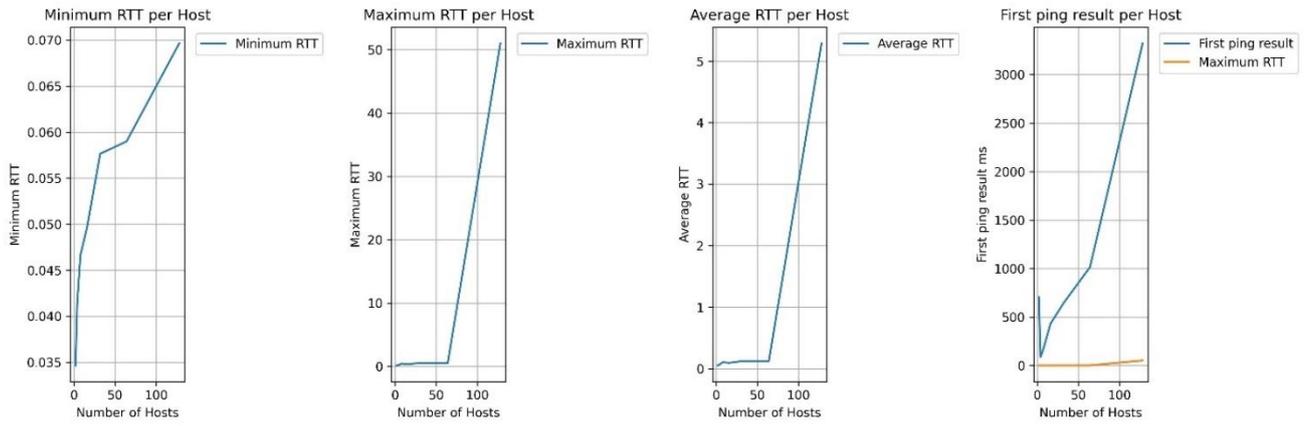

**Figure 21.** RTT Test Results for the Binary Tree Topology

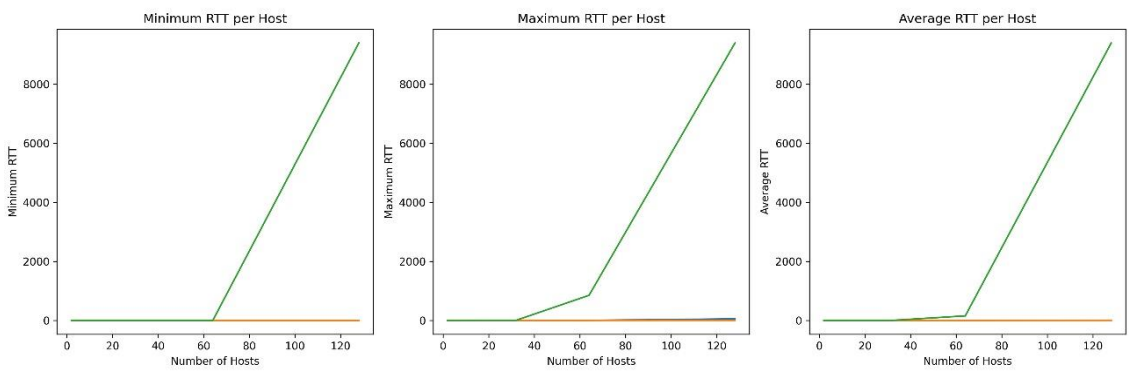

**Figure 22.** RTT Comparison of Loop-Free Topologies

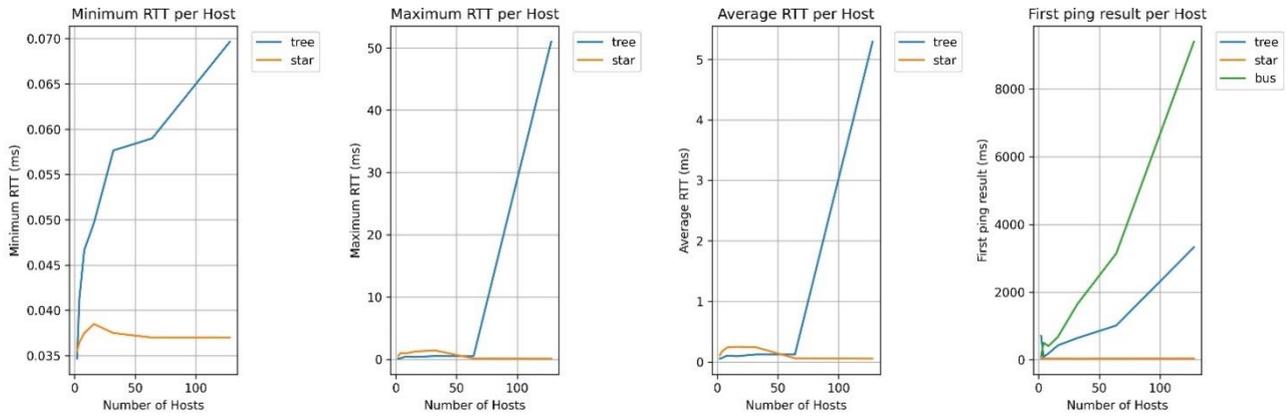

**Figure 23.** RTT Comparison of Star and Binary Tree Topologies

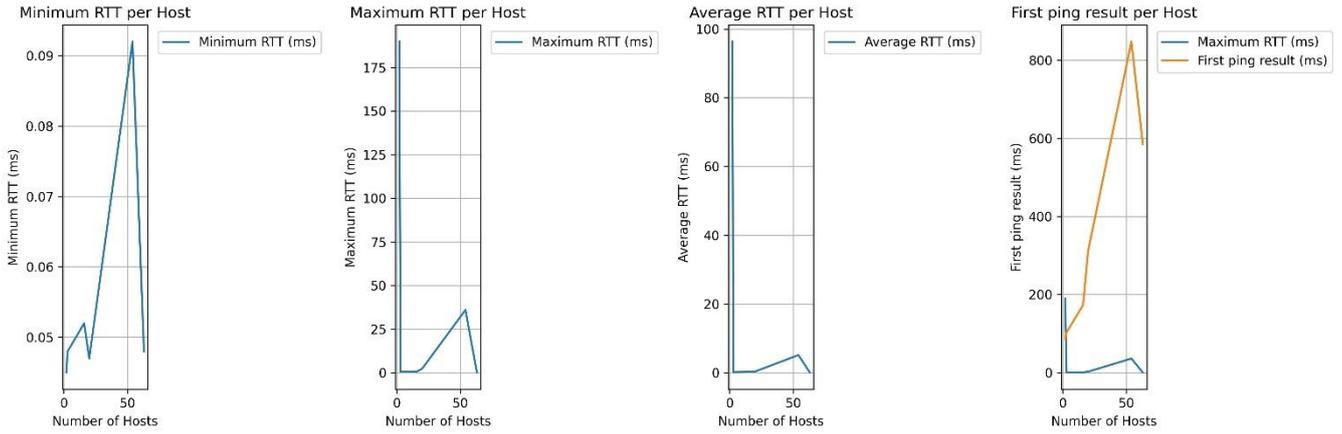

**Figure 24.** **RTT Test Results for the Fat Tree Topology**

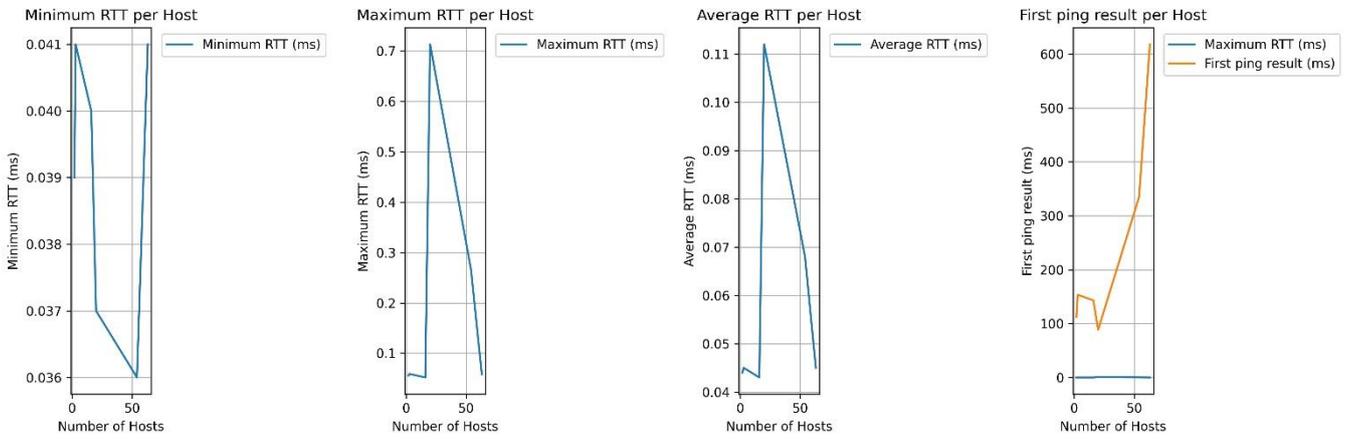

**Figure 25.** **RTT Test Results for the Spine-Leaf Topology**

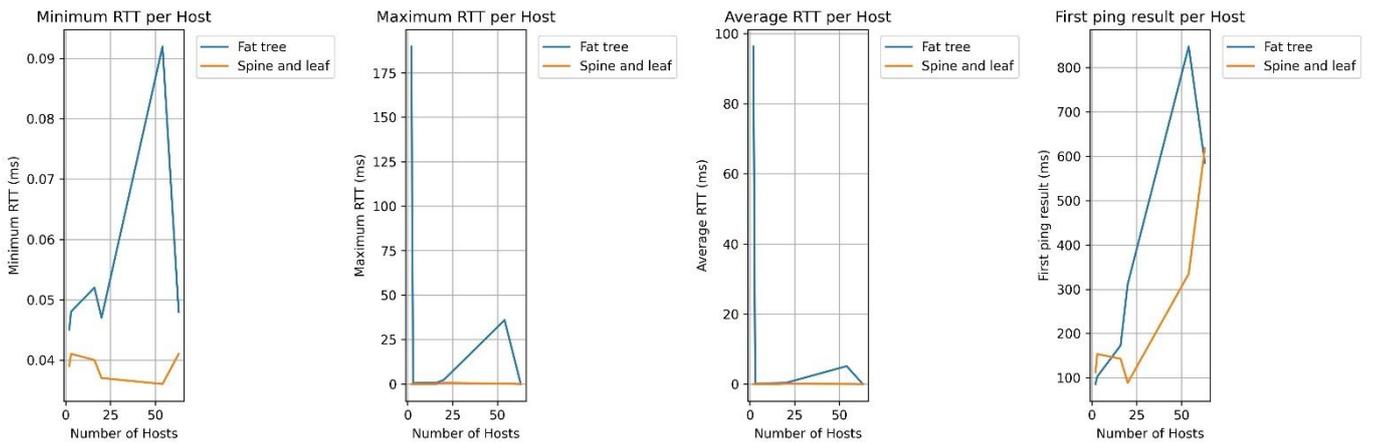

**Figure 26.** **compares the RTT in looped topologies**